\documentclass[10pt,twocolumn,letterpaper]{article}

\usepackage{cvpr}              % To produce the CAMERA-READY version
\usepackage{graphicx}
\usepackage{amsmath}
\usepackage{amssymb}
\usepackage{booktabs}
\usepackage[accsupp]{axessibility}
\usepackage{makecell}
\usepackage{mathtools}
\usepackage{multirow}
\usepackage{hhline}
\usepackage{cellspace}
\usepackage{siunitx}

\usepackage{anyfontsize}
\usepackage{caption}

\usepackage[dvipsnames]{xcolor}

\definecolor{cvprblue}{rgb}{0.21,0.49,0.74}
\usepackage[pagebackref,breaklinks,colorlinks,citecolor=cvprblue]{hyperref}

\usepackage{ulem}

\newcommand{\blue}[1]{\textcolor[rgb]{0,0.4,0.8}}

\renewcommand{\sout}[1]{}

\def\paperID{5} % *** Enter the Paper ID here
\def\confName{CVPR}
\def\confYear{2024}

\title{Unsupervised Microscopy Video Denoising}

\author{Mary Aiyetigbo$^{1}$ \quad Alexander Korte$^{1}$ \quad Ethan Anderson$^{1}$ \quad Reda Chalhoub$^{2}$ \\ \quad Peter Kalivas$^2$ \quad Feng Luo$^{1}$ \quad Nianyi Li$^1$ \vspace{0.3em} \\
{\normalsize $^1$School of Computing, Clemson University} \\ 
\quad
{\normalsize $^2$Department of Neuroscience, The Medical University of South Carolina} \quad \\}

\begin{document}
\maketitle

%%%%%%%%% ABSTRACT
\begin{abstract}

In this paper, we introduce a novel unsupervised network to denoise microscopy videos featured by image sequences captured by a fixed location microscopy camera. Specifically, we propose a DeepTemporal Interpolation method, leveraging a temporal signal filter integrated into the bottom CNN layers, to restore microscopy videos corrupted by unknown noise types. Our unsupervised denoising architecture is distinguished by its ability to adapt to multiple noise conditions without the need for pre-existing noise distribution knowledge, addressing a significant challenge in real-world medical applications. Furthermore, we evaluate our denoising framework using both real microscopy recordings and simulated data, validating our outperforming video denoising performance across a broad spectrum of noise scenarios. Extensive experiments demonstrate that our unsupervised model consistently outperforms state-of-the-art supervised and unsupervised video denoising techniques, proving especially effective for microscopy videos. The project page is available at \href{https://maryaiyetigbo.github.io/UMVD/}{https://maryaiyetigbo.github.io/UMVD/}

\end{abstract}

%%%%%%%%% BODY TEXT

% \begin{figure*}
% \centering
%     \includegraphics[width=0.9\textwidth]{images/teaser_fig.png}
%     % \includegraphics[width=0.9\textwidth]{images/Untitled Diagram.pdf}
    
% \vspace{0.25pt}
% \captionof{figure}{This figure showcases the comparative performance of our method and UDVD on videos contaminated with high-intensity Gaussian, Poisson, and Impulse noise at a high frame rate. Each row illustrates the denoising results for a particular type of noise. From left to right, the frames are arranged in the following order: the original noisy frame, the corresponding clean frame, the denoised result produced by UDVD \cite{udvd}, and finally, the denoised outcome achieved using our method. The presented results clearly indicate our method's robustness and superior performance under different noise types and intensities.}
% \label{fig:teaser}
% \end{figure*}

\section{Introduction}
\label{sec:intro}

In biomedical research, microscopy imaging have enabled scientists to observe and analyze biological structures at cellular and molecular levels \cite{al2020critical, kasban2015comparative, hussain2022modern}. These videos, however, are characteristically captured at fixed locations, limiting the field of view to specific areas of interest within the neural landscape. This fixed location feature, while beneficial for focused studies, introduces unique challenges in video denoising, leading to persistent noise patterns that are difficult to differentiate from actual biological signals \cite{sawinski2009visually, burgess2016hunger, berdyyeva2014zolpidem, pinto2015cell, ziv2013long}. 

In particular, the noise in microscopy videos, emanating from diverse sources such as photon shot noise, background fluorescence, and detector electronics, not only varies in type (Poisson and Gaussian) but also fluctuates in intensity across different frames due to variations in lighting conditions, specimen responses, and camera sensitivity. Traditional video denoising techniques often rely on detecting and leveraging motion between frames to distinguish noise from signal \cite{noise2noise, noise2self, noise2void, deepinterpolation, self-supervised, untrainednet}. 

\begin{figure}[t]
\centering
    \includegraphics[width=1\linewidth]{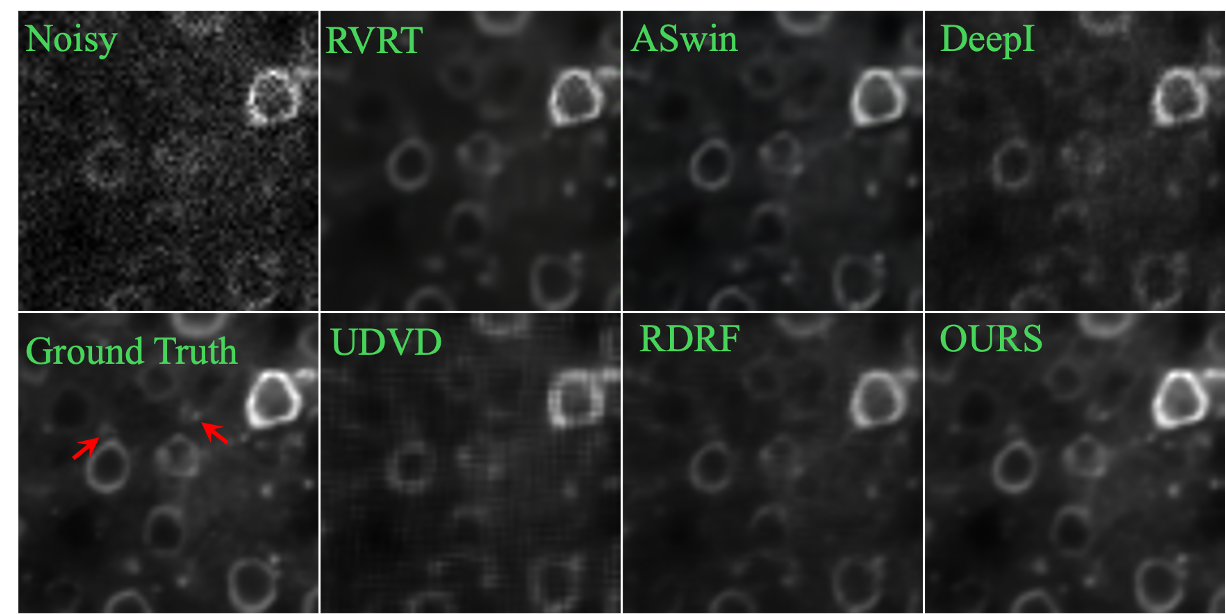}
% \vspace{0.25pt}
\captionof{figure}{Denoising results of our method, supervised (RVRT \cite{rvrt} ASwin \cite{lindner2023lightweight}), and unsupervsied SOTAs (Deepinterpolation (DeepI) \cite{deepinterpolation}, UDVD \cite{udvd}, RDRF \cite{wang2023recurrent}) on simulated two-photon calcium imaging. Our technique showcases superior denoising capabilities. The red arrow indications in the clean image highlight specific signals that were either not adequately reconstructed by other methods in their noise-reduced images or were excessively smoothed}
\label{fig:teaser}
\end{figure}

In the context of microscopy videos with fixed locations, the absence of extensive movement can limit the effectiveness of these approaches, making it difficult to apply conventional denoising techniques designed for more dynamic videos. Furthermore, the varying frame rates in microscopy videos, especially calcium imaging recordings, add an additional layer of difficulty, as they require adaptive processing techniques capable of handling these temporal variations without compromising the integrity of the neural signals being observed.
Therefore, creating effective denoising methods for microscopy videos that can account for static frames while accurately isolating and eliminating noise is significantly important for biological and neurological research.

Many state-of-the-art (SOTA) video denoising methods \cite{deformableattention, fastdvdnet, rvidenet, videnn, dvdnet, rvrt, vrt, RCD, lindner2023lightweight} rely heavily on supervised learning, which require noisy/clean pairs during training. Although they present good denoising performance on videos captured by regular RGB cameras, the supervised video denoising methods show limited generalization ability in medical imaging, where obtaining clean reference images is unrealistic and almost impossible.
Various unsupervised approaches address this challenge by optimizing the noisy frames and exploring the motion between frames \cite{noise2noise, noise2self, noise2void, deepinterpolation, self-supervised, untrainednet}. Nevertheless, SOTA unsupervised video denoising networks, \ie UDVD \cite{udvd}, Wang \etal \cite{wang2023recurrent}, have inherent limitations. Firstly, they employ the backbone network architectures of the Multi-Frame2Frame \cite{frame2frame,fastdvdnet} and Laine \etal \cite{self-supervised}, which extensively exploits information from neighboring pixels, both spatially and temporally, to reconstruct the missing data and eliminate noise. These approaches prove effective primarily for low frame rate videos or videos featuring fast-moving objects or view changes, and easy to overfit to noise when the input videos exhibit slow-moving objects with consistent backgrounds. However, it is common for microscopy videos, \eg, one-photon calcium imaging, containing static but fluctuating noisy background with only a few intermediate signal spikes throughout the recording, thus can pose extreme challenges to these motion-based denoising techniques, as shown in Fig.~\ref{fig:teaser}.  

% This is primarily caused by the strong resemblance in structural information among neighboring frames. 
% As displayed in Fig.~\ref{fig:teaser}, UDVD demonstrates degraded performance in these scenarios, signifying the need for a more versatile denoising approach to cater to a broader range of video content and frame rates.
% Moreover, many CNN-based denoising models (both supervised and unsupervised techniques) are primarily developed to remove Gaussian noise. Their performance tends to degrade when handling other types of noise, including Poisson and Impulse noise , highlighting a significant limitation in the existing approaches.

\begin{figure*}
\centering
    \includegraphics[width=1\textwidth]{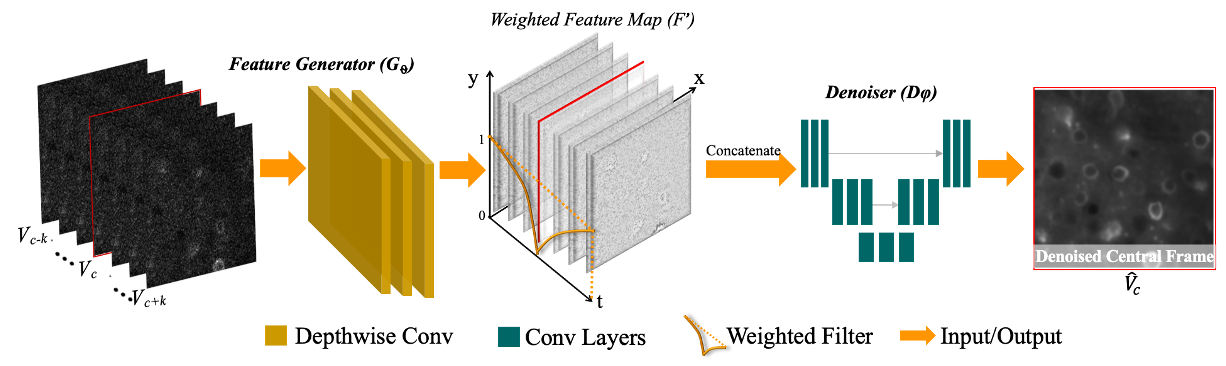}
% \vspace{-10pt}
\captionof{figure}{Our DeepTemporal Interpolation Pipeline. The feature generator $\mathbf{F}_{\phi}$ extracts distinctive features using three depthwise convolutional layers, and a temporal filter enhances denoising accuracy by adjusting feature map weights based on spatial-temporal proximity to the central frame, overcoming challenges in handling high frame rate videos and slow-moving objects.}
\label{fig:pipeline}
\end{figure*}

In this paper, we present an unsupervised denoising method using DeepTemporal Interpolation, tailored for microscopy videos, which can effectively denoise videos with varied noise types and intensities. Our method consists of two main components: a feature generator ($\mathcal{G}_\phi$) and a Denoiser ($\mathcal{D}_\theta$), as shown in Fig.~\ref{fig:pipeline}. Distinct from conventional unsupervised methods, which directly use adjacent frames to interpolate the missing central frame, such as DeepInterpolation \cite{deepinterpolation} and Zheng \etal \cite{untrainednet}, or utilize neighboring pixels for central pixel estimation as seen in UDVD \cite{udvd} and RDRF\cite{wang2023recurrent}, our technique employs a temporal filter on the feature maps produced by a sequence of CNN layers ($\mathcal{G}_\phi$) prior to processing by the Denoiser ($\mathcal{D}_\theta$). 
Our overall pipeline is illustrated in Fig.~\ref{fig:pipeline}. The major contributions of our work include:

\begin{itemize}

    \item We propose a novel unsupervised video denoising approach tailored for microscopy videos. Instead of making complex modifications to the network architecture, we apply a temporal filter to the feature map generated by the feature generator which is fed into the denoiser, effectively mitigating the overfitting challenge commonly encountered in unsupervised models.

    \item Through comprehensive evaluations using both real and simulated medical imaging datasets, our model outperforms supervised and unsupervised denoising methods.

    \item Our technique also represents a significant advancement in denoising videos containing non-Gaussian noise types, an area where many existing CNN-based models struggle.
    
    % \item Our technique represents a significant advancement in handling non-Gaussian noise types, an area where many existing CNN-based models struggle. This enhancement in adaptability and performance increases the relevance of our model in various real-world scenarios and natural color video datasets.
\end{itemize}

\section{Related Work}

% In this section, we review the relevant literature on both supervised and unsupervised video denoising. Note that we do not survey the large literature on image denoising algorithms~\cite{xu2015patch, zhang2011learning, lebrun2013nonlocal, dncnn, noise2noise, mao2016image} as our method is primarily concerned with video denoising techniques.

\paragraph{Supervised Video Denoising Networks}

Supervised deep learning-based approaches \cite{dncnn, noise2noise, mao2016image, fastdvdnet, frame2frame} have demonstrated superior performance in video denoising. 
% For instance, VNLnet \cite{VNLnet} utilizes non-local search to pinpoint similar video patches for CNN processing. 
DVDNet \cite{dvdnet} leverages UNet \cite{unet} blocks to process five successive neighboring frames, aiming to denoise the central frame with optical flow to compensate for motion explicitly. FastDVDnet \cite{fastdvdnet} improved on \cite{dvdnet} and performs implicit motion compensation through the architecture design. Vision Transformer-based methodologies such as those presented in VRT \cite{vrt}, RVRT \cite{rvrt}, ASwin \cite{lindner2023lightweight} and Song \etal \cite{song2022tempformer} leverage transformer architectures to exploit temporal information in the video, thereby enhancing their noise reduction efficiency. A notable limitation in these supervised techniques is the need for training datasets comprising noisy and clean pairs, which may not always be feasible in real-world settings. Moreover, these models are generally tailored for specific noise types. This potentially constrains their versatility when confronted with different noise conditions, resulting in diminished performance on noise patterns not encountered during training.

\paragraph{Unsupervised Video Denoising Networks}
Unsupervised video denoising methods mitigate the challenge of requiring noisy/clean video pairs.
Noise2Noise \cite{noise2noise} demonstrates comparable image denoising performance with supervised one by learning the underlying clean image from pairs of noisy images. Frame2Frame \cite{frame2frame} extended this concept to video denoising by treating sequential frames as pairs of noisy images and employing optical flow techniques to estimate motion. However, the performance of this method can be limited by inaccuracies in motion estimation. Dewil \etal \cite{mf2f} extends single-image approach in \cite{frame2frame} by introducing a self-supervised fine-tuning framework which adopts FastDVDnet \cite{fastdvdnet} as the backbone network.
The Noise2Void method \cite{noise2void} introduced a 'blind-spot' technique which estimates each noisy pixel by considering neighboring pixels without including the noisy pixel itself. Laine \etal \cite{self-supervised} enforced the blind-spot concept through architectural design.
The state-of-the-art (SOTA) unsupervised video denoising method, \ie UDVD \cite{udvd}, integrated the blind-spot network (BSN) in \cite{self-supervised} into the FastDVDnet \cite{fastdvdnet} framework. Wang \etal \cite{wang2023recurrent} increases the receptive field of the BSN and also includes a transformer network to capture temporal information. These methods have the limitation of overfitting noise when neighboring frames have very similar structures and require complex training. Zheng \etal \cite{untrainednet} developed an unsupervised loss function during training to provide an unbiased estimator. However, this approach requires prior knowledge of noise distribution, which might not be available when denoising real videos.

The most related to our approach is DeepInterpolation \cite{deepinterpolation}, which uses U-Net as a backbone network. It denoises videos by omitting the central frame and leverages neighboring frames' similarity for interpolation. However, this method struggles when handling scenarios with high noise intensity. In contrast, our innovative model seamlessly integrates a temporal filter into the feature map generated by stacked CNN layers. 
By allocating higher weights to distant frame feature maps and lower weights to proximate ones, our model addresses the rapid convergence to noisy content seen in high frame rate videos due to redundancy in nearby frames, especially when the noise level is high. Consequently, our method showcases more resilient denoising performance than DeepInterpolation and other unsupervised approaches in UDVD and RDRF.

\section{Unsupervised Video Denoising}
Our goal is to generate denoised video $\{\hat{\mathbf{V}}_t|t = 1,2,...T\}$ given a sequence of noisy video input, $\{\mathbf{V}_t | t = 1,2,...T\}$, where $T$ is the total number of frames in the input video. In each iteration, a subset of contiguous frames $\{\mathbf{V}_t | t = 1,...c,...N\}$ is taken as input, with $N$ indicating the batch frame count and $c$ denoting the central frame's index. These frames are then passed through the feature generator $\mathbf{G}_{\phi}$, producing the corresponding feature maps $\{\mathbf{F}_t | t = 1,...,N\}$. Then, the temporal filter $\{ \gamma_t\}_{t=1}^N$ weights these feature maps $\{\mathbf{F}_t\}_{t=1}^N$, assigning diminished values to features nearer the central frame compared to those more distant. Next, the resulting weighted feature maps are concatenated and fed into the Denoiser $\mathbf{D}_{\theta}$, producing the denoised central frame $\hat{\mathbf{V}}_c$. A comprehensive visualization of our pipeline is provided in  Fig.~\ref{fig:pipeline}.

\subsection{Feature Generator $G_{\phi}$}
% To effectively denoise a specific frame, our approach processes a batch of consecutive frames, denoted as $\{\mathbf{V}_t | t = 1,2,...c,...,N-1,N\}$, where $N$ is the total number of frames in the batch, $t$ is the index of each frame in the batch and $c$ is the index of the central frame. (\nianyi{Remove all the description of $K$}). 
Given a batched frames $\{{\mathbf{V}}_t\}_{t=1}^N$, the feature generator $\mathbf{G}_{\phi}$ produces feature maps $\{\mathbf{F}_t\}_{t=1}^N$ aligned with the input frames. We utilize depth-wise convolution layers to expedite the training phase, eliminating the need to process each frame individually through $\mathbf{G}_{\phi}$.
Unlike standard 2D convolutions that apply a multi-channel filter to the entire input depth, allowing for channel mixing, depthwise convolution maintains the separation of input channels and generates the output feature maps independently. 
Specifically, our feature generator integrates three depth-wise convolution layers. Each depthwise group generates feature map $\mathbf{F}_t\in \mathbb{R}^{C\times H \times W}$ for $\mathbf{V}_t$, where $C$ is the number of output channels, while $H$ and $W$ represent the input image's height and width, respectively.

\subsection{DeepTemporal Interpolation}
% \begin{figure}
% \centering
%     \includegraphics[width=1\linewidth]{images/teaser_illu.png}
% \captionof{figure}{The illustration of our SpatioTemporal Weighted Kernel.}
% \label{fig:filter_output}
% \end{figure}

% Upon the generation of the feature maps $\mathbf{F}_t$ at the initial stage, we introduce the Temporal filter. This filter plays a crucial role in enhancing denoising accuracy. The temporal filter is utilized as a set of weighted parameters applied to the feature maps. Unlike DeepInterpolation \cite{deepinterpolation}, we do not need to remove the central frame from the input. Instead, we apply the temporal filter, which specifically sets the values of the central frame's feature map to zero while diminishing weights are assigned to the feature maps of the close neighboring frames. This strategic assignment of weights prevents the model from mapping to the inherent noise patterns present in the target and surrounding frames. As a result, the model is better positioned to precisely approximate the underlying clean signal.

% Our temporal filter is distinct in its ability to adjust the weight of each input frame based on its spatial-temporal proximity. This unique weighting allows our approach to manage the intricate spatio-temporal correlations within high frame rate and slow-motion videos more effectively, addressing the shortcomings of existing models. In this vein, the temporal filter offers a new paradigm for video denoising, overcoming the limitations of current methods in handling high frame rate videos and slow-moving objects.

Upon generating the feature maps $\mathbf{F}_t$ at the initial stage, we introduce the DeepTemporal Interpolation method that applies a set of weighted parameters to the feature maps. This filter plays a crucial role in enhancing denoising accuracy. Unlike DeepInterpolation \cite{deepinterpolation}, we do not need to remove the central frame from the input. Instead, our temporal filter is distinct in its ability to adjust the weight of each feature map based on its temporal proximity by setting the values of the central frame's feature map to zero, while diminishing weights are assigned to the feature maps of close neighboring frames. This unique weighting allows our approach to manage the intricate temporal similarities within high frame rate and slow-motion videos more effectively, addressing the shortcomings of existing models. This strategy prevents the model from mapping the inherent noise patterns in the target and surrounding frames. Notably, the temporal filter offers a new paradigm for video denoising, overcoming the limitations of current methods in handling high frame rate videos and slow-moving objects, which is common in microscopy videos.
The temporal filter weights the feature map using Eq \ref{eq:filter}.

% \begin{equation}
% \gamma_t=
% \begin{cases}
% 1-\left(\frac{t}{c}\right)^L & \text{if } t \leq c, \\
% 1-\left(\frac{2c-t}{c}\right)^L & \text{otherwise}
% \end{cases}
% \label{eq:TF}
% \end{equation}

% Here, $\gamma_t$ is the weight of the feature map $\mathbf{F}_t$, $t$ is the index of the frame, $c$ is the index of the central frame $\mathbf{V}_c$, and $L$ is a parameter controlling the curvature of the TF function, as illustrated in Fig.~\ref{fig:TF}.

% \mary{Should we use this linear filter as the temporal filter so that it doesn't look similar to our ECCV submission? What do you think?}

% \nianyi{Yes, please directly use the linear filter.}

\vspace{-5pt}
\begin{equation}
\gamma(.) = \left[\left\{\frac{k-i}{k}\right\}_{i=0}^{k-1}, \left\{\frac{i}{k}\right\}_{i=1}^k\right],
\label{eq:filter}
\end{equation}
where $k=\lfloor{M/2}\rfloor$, and $\lfloor{.}\rfloor$ denotes the floor function.

% \begin{figure}[h]
% \centering
%     \includegraphics[width=1\linewidth]{images/TF_illu.png}
% \captionof{figure}{Temporal Filter (TF) function, which illustrates the weighting scheme. Each curve delineates the TF function corresponding to a distinct curvature parameter value denoted as $L$. The x-axis represents the index of each feature map $t$, wherein the central frame's feature map $\mathbf{F}_c$ is positioned at index $c$. The y-axis signifies the weight $\gamma_t$ assigned to each feature map produced by the feature generator. As shown, varying $L$ influences the distribution of weights across the sequence, with larger $L$ values leading to a steeper descent from the central element.}
% \label{fig:TF}
% \end{figure}

% In terms of preprocessing, we employ a sliding window strategy to concatenate neighboring frames into a large 3D tensor $\{\mathbf{G}_i|i=1,...,M\}$ (with $M=7$ in our setting). This 3D tensor then forms the input for a U-Net architecture that predicts the central frames. Notably, due to the U-Net's inherent skip connections, direct input of this 3D tensor could lead to an undesirable identical mapping of the noisy central frame to the output. To avoid this, we apply the weights $\{\gamma_i|i=1,...,M\}$ to all the frames, ensuring the weight of the central frame $\mathbf{G}_C$ is zero, and the weights of the first and last frames ($\mathbf{G}_1$ and $\mathbf{G}_M$) are one. This process can be mathematically represented as:

We apply the weights $\{\gamma_t|t=1,...,N\}$ to the feature maps by performing an elementwise multiplication. This ensures that the weight of feature map $\mathbf{F}_c$ at index $c$ is zero, and the weights of the first and last feature map in the batch ($\mathbf{F}_1$ and $\mathbf{F}_N$) are one. This operation is depicted in Eq \ref{eq:weighted}.

\vspace{-5pt}
\begin{equation}
\mathbf{F}'_t = \gamma_t \odot \mathbf{F}_t \quad \forall t \in {1,...,N}.
\label{eq:weighted}
\end{equation}

Through this innovative application of the temporal filter, our method is able to provide robust denoising for high frame rate and slow-motion videos, effectively overcoming the challenges that limit current state-of-the-art approaches.

\subsection{Denoiser $D_{\theta}$}
Given the weighted feature maps $\{\mathbf{F}'_t | t = 1,...,c,...,N\}$, the Denoiser $D_{\theta}$ generates a denoised central frame $\hat{\mathbf{V}}_c$ as in Eq \ref{eq:denoiser}.
The Denoiser $D_{\theta}$ model is a U-Net \cite{unet} architecture with three encoders and two decoder layers.
The concatenation of $\{\mathbf{F}'_t\}_{t=1}^N$ serves as the input to $D_{\theta}$, enabling it to explore temporal correlations within adjacent frames. 

\vspace{-5pt}
\begin{equation}
\hat{\mathbf{V}}_c = \mathcal{D}_\theta([\mathbf{F}'_1,...\mathbf{F}'_N]).
\label{eq:denoiser}
\end{equation}

All convolutional layers in the network utilize ReLU activation functions except the output layer. It is important to note that the integration of skip connections between input and output layers in U-Net inherently amplifies the network's sensitivity to high-frequency details present in the input tensor, encompassing the inherent noise. Consequently, assigning a value of $\gamma_c=0$ is imperative to guarantee the exclusion of any original central frame information from being transferred to $D_{\theta}$. We found that even marginal values for $\gamma_c$ can prompt the U-Net to rapidly converge towards the noise during training.

\section{Training Details.}

In our implementation, we take a stack of $N=7$ contiguous frames. 
We adopt training protocols consistent with other unsupervised image and video denoising techniques, such as Laine \etal \cite{self-supervised} and UDVD \cite{udvd}. Specifically, the input tensor is partitioned into overlapping patches of size $128\times128$ in spatial dimensions, from which we predict the denoised central patches. This procedure, which is a form of data augmentation, significantly augments the number of training samples, thereby enhancing the denoising performance.
It is essential to note that while we utilize these cropped patches during training, the full-resolution tensor was used for denoising during the inference phase.
% We follow the training practices in other unsupervised image and video denoising methods, like Laine \etal \cite{self-supervised} and UDVD \cite{udvd}, which cropped the input large tensor to a sequence of overlapping smaller patches of size $128\times128$ along spatial dimensions, and predict the noisy-free central ones. This step, belonging to data augmentation, can greatly increase the training samples and increase the noising performance. 

% In line with established practices in image and video denoising, our network takes as input stacks of $N=7$ patches, cropped from contiguous noisy frames $\mathbf{V}_t$ at the same spatial location. The patches are of size $128\times128$ and are derived from our dataset. As part of our training data augmentation strategy, we randomly crop these patches from the input tensor along its spatial dimensions.

% For natural videos, we create a synthetic noise version for each frame, save the resultant noisy frames, and load them into our model as inputs. This ensures a consistent noise value across all frames throughout the training process.
We employ the $l_2$ loss function in Eq \ref{eq:loss} to minimize the difference between the model's output $\hat{\mathbf{V}}_c$ and the reference central noisy frame $\mathbf{V}_C$

\vspace{-5pt}
\begin{equation}
    \mathcal{L} = ||\hat{\mathbf{V}}_c-\mathbf{V}_c||^2_2.
    \label{eq:loss}
\end{equation}

% It's worth mentioning that, while we use these cropped patches for training, during the inference phase, we use the entire resolution tensor as input to our model to denoise the video content.

\section{Experiments}
We carried out a series of evaluations to assess the efficiency of our method. Our results show that our approach surpasses many SOTA video denoising techniques, including both supervised and unsupervised methods. Additionally, our ablation studies offer valuable understanding regarding the importance of key components within our methodology.

\subsection{Datasets}
We evaluated our method across diverse datasets, comprising both real noisy sequences in the case of calcium imaging and those infused with synthetic noise. For the natural video, sequences were contaminated with a range of noise types and intensities to simulate various challenging conditions. For medical imaging, we used real-world noisy videos primarily from fluorescence microscopy and calcium imaging. We used the recordings from neural recordings of freely moving mice for one-photon imaging. To evaluate with two-photon imaging, we simulated realistic calcium imaging, which also has the ground truth clean imaging.
% \nianyi{Need to add the microscopy dataset in the Supplement materials.}

\paragraph{Real-world Dataset}
Our real-world dataset comprises one-photon calcium imaging recordings procured locally from freely behaving transgenic mice (Drd1-Cre and Drd2-Cre) during cocaine/sucrose self-administration experiments. This calcium imaging technique, critical in neuroscience, allows for real-time visualization of neuronal activity. However, these recordings are often degraded by noise, making denoising an essential preprocessing step.
For these recordings, single-channel epifluorescent miniaturized miniscopes were employed, capturing the fine details of the neuronal activity. Given that this dataset possesses complex noise levels and lacks reference ground-truth clean videos, it provides a valuable practical demonstration of our model's application to real-world scenarios.

For fluorescence microscopy, we utilized the microscopy recordings of live cells obtained from \cite{ulman2017objective}. We used the GOWT1 cell recording (Fluo-C2DL-MSC) and mesenchymal stem cell recording video (Fluo-N2DH-GOWT1). Similar to the one-photon calcium imaging, this data has no ground truth clean video.

We simulated realistic two-photon calcium imaging data using neural anatomy and optical microscopy (NAOMi) \cite{naomi} to perform the qualitative evaluation on realistic imaging with ground truth. We generated two datasets with 1,000 and 1,500 frames, respectively, with the field of view (FOV) of $150$x\SI{150}{\micro\meter}$^2$ and $500$x\SI{500}{\micro\meter}$^2$ respectively.

\paragraph{High Frame Rate RGB Video Dataset.}
% \nianyi{Need to briefly explain why we need these color video datasets for testing. Why not directly using  NaoMi to synthesize the dataset?}
To validate our model's generalization capabilities across a broad spectrum of video content, we extend our evaluation beyond microscopy imaging to include color video data captured with RGB cameras. This approach ensures a comprehensive evaluation under diverse noise types and intensities, showcasing the model's adaptability to real-world applications. To achieve this, we corrupted the video sequence with synthetic noise, employing three noise types: Gaussian, Poisson, and Impulse (Salt-and-Pepper).
We used the LIVE YouTube High Frame Rate (LIVE-YT-HFR) \cite{livehfr} datasets to evaluate on high frame rate RGB video. This dataset features a wide array of video sequences, each captured at various frame rates per second (fps) and subject to different compression levels. Specifically, our evaluation concentrated on sequences captured at 120fps. This demonstrates the robustness of our model across various high frame rate video scenarios, further affirming its efficacy and wide-ranging applicability in video denoising tasks.

To simulate varying intensities of Gaussian noise, we used standard deviation values ($\sigma$) of 30, 50, and 90. Intensities of Poisson noise (photon shot noise) were varied by setting the maximum event count ($\lambda$) to 30, 50, and 90. For Impulse noise, which appears as random white or black pixels due to sudden signal changes, we used pixel ratios ($\alpha$) of 0.2, 0.3, and 0.4 to indicate different noise levels.

\subsection{Experimental Setup}
We benchmarked our DeepTemporal Interpolation against SOTA unsupervised video denoising techniques DeepInterpolation \cite{deepinterpolation}, UDVD \cite{udvd}, RDRF \cite{wang2023recurrent}, and also against the supervised paradigm, RVRT \cite{rvrt}, and ASwin \cite{lindner2023lightweight}.
All models were implemented using the PyTorch framework \cite{pytorch} and NVIDIA A100 GPU was used for execution. We adopted the ADAM optimizer with default parameters. Over a span of 25 epochs, each iteration employed an input stack as the mini-batch, targeting the central frame's denoising. An early stopping strategy was used to mitigate potential overfitting, halting the training process in the absence of notable loss decrement. The inception learning rate was fixed at $1e-3$, and was systematically halved every 10 epochs.

For the supervised video denoising methods, \ie RVRT \cite{vrt} and ASwin \cite{lindner2023lightweight}, we use their publicly available pre-trained model, which was already trained under various noise conditions to generate the denoising results. Specifically, for ASwin, we used the model trained for blind denoising of real-world noisy video. For UDVD \cite{udvd}, we train individual video sequences using the \textbf{UDVD-S} model. The training process for \textbf{UDVD-S} followed a blind denoising approach, minimizing the mean squared error (MSE) without any assumptions regarding the noise parameters. Similar to UDVD, we also trained each video sequence separately for DeepInterpolation \cite{deepinterpolation}. For RDRF \cite{wang2023recurrent}, we generated the denoising result for the color video data using the pre-trained model provided by the authors. We specifically pass the noisy input to the model without giving any prior information about the noise parameters. However, we trained the model from scratch on each video recording for the microscopy dataset. All implemented models were trained and evaluated on the same datasets, ensuring consistency in our comparative analysis.

\begin{figure*}[htbp]
\centering
    \includegraphics[width=.99\textwidth]{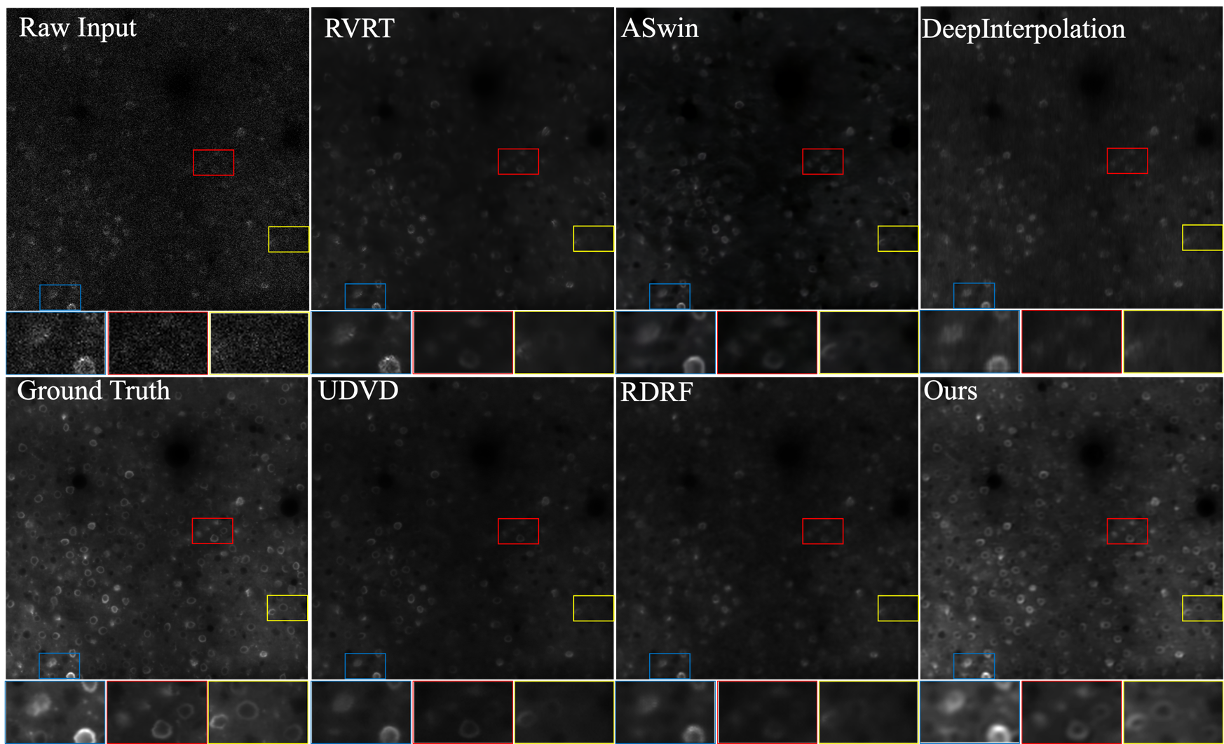}

\vspace{-5pt}
\captionof{figure}{This figure presents a comparative analysis of denoising performance between our approach and that of other established methods, including RVRT, Aswin, DeepInterpolation, UDVD, and RDRF, on Two-Photon (2P) calcium imaging data simulated using NAOMi to produce realistic imaging with a ground truth reference. The outcomes illustrate that our model successfully denoises the video data while preserving relevant signals and avoiding obscuring critical information. Conversely, the denoising outputs from alternative methods reveal their limitations in accurately recovering underlying signals. }
% \nianyi{You need to add bounding boxes to ALL images, instead of only the input.}}
\label{fig:visual_cmp}
\end{figure*}

It's worth noting that while specific unsupervised image/video denoising methodologies, including Laine \etal \cite{self-supervised}, UDVD \cite{udvd} RDRF \cite{wang2023recurrent}, Zheng \etal \cite{untrainednet} boost their efficacy by introducing random noise to clean reference frames in the training phase, we deliberately refrained from employing such strategies. Utilizing these techniques can result in overfitting to a particular noise characteristic, often Gaussian, and a specific noise intensity. This, in turn, degrades the model's adaptability to real-world, varied noise situations. Consequently, our training strategy only utilizes the inherent noisy video frames as input, avoiding the addition of any noise during training.

\begin{figure*}[htbp]
\centering
    \includegraphics[width=.99\textwidth]{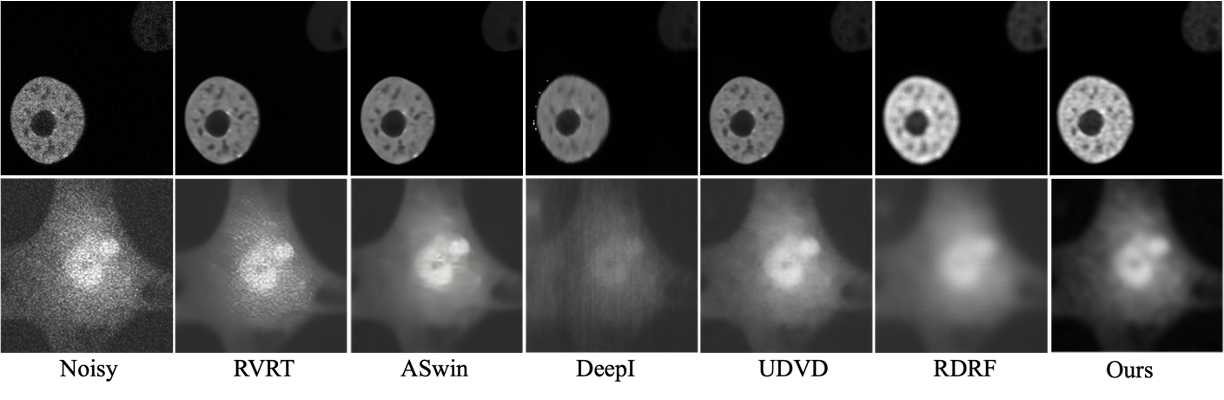}

\vspace{-10pt}
\captionof{figure}{Visual Comparison on Flourescence Microscopy}
\label{fig:visual_floro}
\end{figure*}

\begin{figure*}[htbp]
\centering
    \includegraphics[width=1\textwidth]{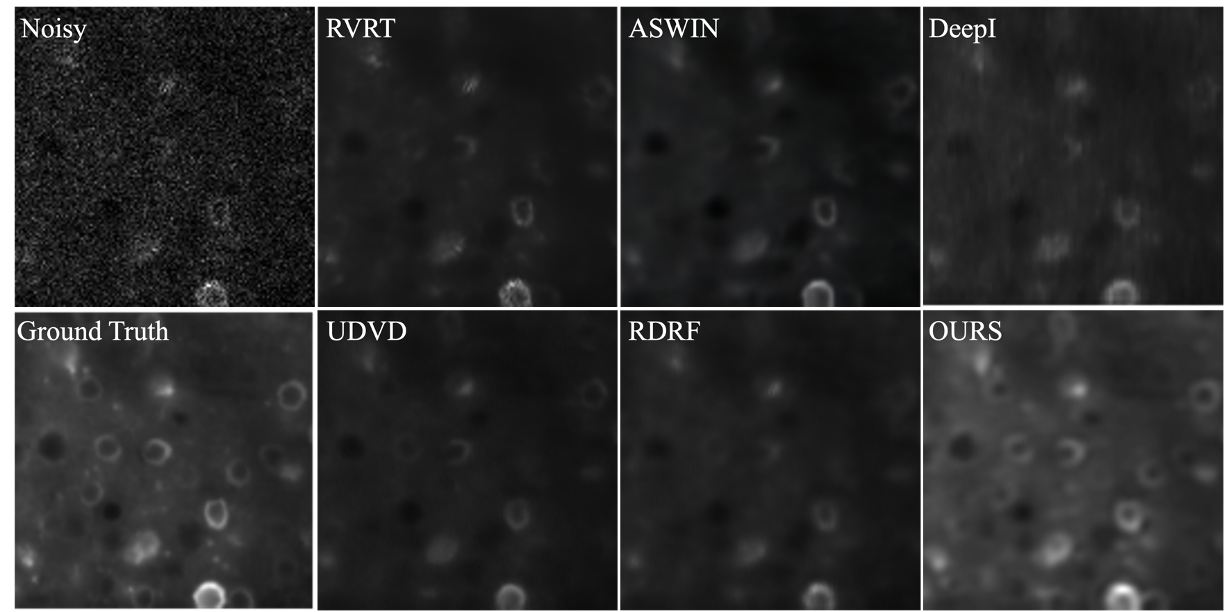}

\vspace{-5pt}
\captionof{figure}{Additional visual comparison of two-photon calcium imaging}
\label{fig:visual_2pcal}
\end{figure*}

\begin{figure*}[htbp]
\centering
    \includegraphics[width=1\textwidth]{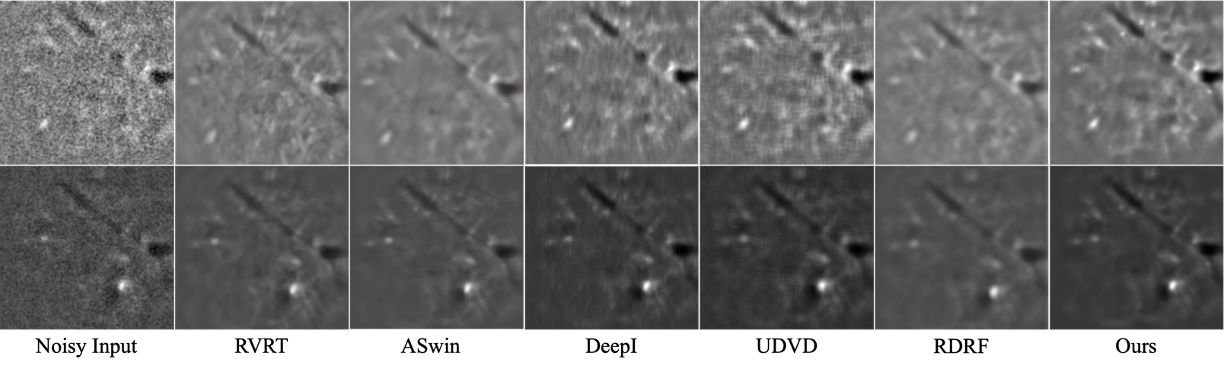}

\vspace{-5pt}
\captionof{figure}{{Denoising results on one-photon calcium imaging recordings of freely behaving transgenic mice}}
\label{fig:musc}
\end{figure*}

\newcommand{\Gauss}{\multicolumn{1}{c}{$\sigma=30$} & \multicolumn{1}{c}{$\sigma=50$} & \multicolumn{1}{c}{$\sigma=90$}}
\newcommand{\Pois}{\multicolumn{1}{c}{$\lambda=30$} & \multicolumn{1}{c}{$\lambda=50$} & \multicolumn{1}{c}{$\lambda=90$}}
\newcommand{\Imp}{\multicolumn{1}{c}{$\alpha=0.2$} & \multicolumn{1}{c}{$\alpha=0.3$} & \multicolumn{1}{c}{$\alpha=0.4$}}
\begin{table*}[t]
% \resizebox{\textwidth}{}{%
% \fontsize{3pt}{3pt}{3pt}{3pt}{3pt}{3pt}{3pt}{3pt}{3pt}{3pt}{3pt}\selectfont
\small
  \centering
  \setlength{\abovetopsep}{7pt}
  \setlength{\tabcolsep}{4.5pt}
  \renewcommand{\arraystretch}{1}
  \begin{tabular}{*{11}{l}}
    \toprule
    & & \multicolumn{3}{c}{Gaussian} & \multicolumn{3}{c}{Poisson}   & \multicolumn{3}{c}{Impulse}  \\
    \cmidrule(lr){3-5}
    \cmidrule(lr){6-8}
    \cmidrule(lr){9-11}
& & \Gauss & \Pois  &\Imp \\
    \midrule
    \multirow{4}{*}{\parbox{1.2cm}{LIVE-YT-HFR 120fps}}
        & RVRT &  \underline{30.75}/{0.80} & \textbf{29.61}/\underline{0.83} & {18.12}/{0.22} & {27.68}/{0.78} & \underline{27.27}/\textbf{0.92} & {17.64}/{0.22} & {23.84}/{0.63} & {17.32}/{0.19} & {13.98}/{0.09}\\[1pt]

        & ASwin &  {15.50}/{0.40} & {15.55}/{0.41} & {15.90}/{0.42} & {15.12}/{0.39} & {14.86}/{0.39} & {14.29}/{0.39} & {15.81}/{0.42} & {16.10}/{0.42} & {16.18}/{0.42}\\[1pt]
        
        & DeepI & {30.25}/\underline{0.87} & {28.18}/{0.82} & \textbf{22.94}/0.71 & {28.09}/\underline{0.89} & {26.36}/{0.88} & \underline{23.25}/\underline{0.86} & \underline{23.86/0.75} & \underline{21.15}/\underline{0.67} & \underline{19.06}/\underline{0.61} \\[1pt]
 
        & UDVD & {28.47}/{0.84} & {25.66}/{0.76} & {22.15}/{0.65} & \underline{29.37}/{0.87} & {26.24}/{0.84} & {22.18}/{0.74} & {22.48}/{0.68} & {20.24}/{0.61} & {18.45}/{0.57} \\[1pt]
        
        & RDRF &  {25.80}/{0.58} & {20.08}/{0.35} & {15.35}/{0.18} & {25.43}/{0.65} & {22.39}/{0.53} & {18.99}/{0.41} & {16.79}/{0.23} & {14.61}/{0.16} & {13.14}/{0.12}\\[1pt]

         & Ours & \textbf{31.67}/\textbf{0.90} & \underline{28.70}/\textbf{0.85} & \underline{22.82}/\textbf{0.75} & \textbf{30.82}/\textbf{0.92} & \textbf{28.72}/\underline{0.89} & \textbf{24.76}/\textbf{0.88} & \textbf{24.44}/\textbf{0.82} & \textbf{21.29}/\textbf{0.74} & \textbf{19.23}/\textbf{0.67} \\
        
        \midrule
    \bottomrule
  \end{tabular}
  \vspace{-10pt}
\caption{\textbf{Performance in Denoising Synthetic Noise.} This table presents a comparison of average PSNR/SSIM values of denoised performance on LIVE-YT-HFR datasets. Text highlighted in \textbf{bold} signifies the highest value, while \underline{underlined} text denotes the second highest. Our method demonstrates superior performance in most cases and remains highly competitive with the supervised methods.}
\label{tab:resultstab}
\end{table*}

\begin{table}[h]
\small
  \centering
  \setlength{\abovetopsep}{7pt}
  \setlength{\tabcolsep}{3pt}
  \renewcommand{\arraystretch}{1.2}
  \begin{tabular}{cccccccc}
  \hline
  \centering
   &  & RVRT & ASwin & DeepI & UDVD & RDRF & Ours\\
   \hline
   & PSNR & \underline{30.05} & {25.94} & {28.64} & {25.75} & {28.31} & \textbf{35.90}\\
   & SSIM & \underline{0.92} & {0.90} & {0.91} & {0.74} & \underline{0.92} & \textbf{0.95}\\
   \hline
  \end{tabular}
    \vspace{-10pt}
  \caption{\textbf{Denoising performance on simulated 2P Calcium Imaging.} The result shows the average PSNR and SSIM values on the 2P imaging simulated with NAOMi \cite{naomi}}
  \label{tab:snr}
\end{table}

\subsection{Quantitative Evaluation}
To evaluate our proposed method, we used two metrics in the domain of image and video denoising: the Peak Signal-to-Noise Ratio (PSNR) and Structural Similarity Index Measure (SSIM)
% Our choice to utilize these metrics aligns with the conventional experimental setup employed for comparative analyses in the field.
As shown in Table \ref{tab:resultstab}, our method outperforms the unsupervised methods, RDRF and UDVD, across various noise categories on the RGB video dataset. While the supervised RVRT and ASwin models demonstrate good denoising performance on Gaussian noise, their performances significantly drop relatively on other noise types, especially in scenarios with high noise levels. Additionally, the findings in Table \ref{tab:snr} demonstrate that our method surpasses supervised and unsupervised techniques on the simulated two-photon calcium imaging dataset. This corroborates our model's efficacy in denoising across various datasets and noise scenarios, showcasing its broad applicability and effectiveness.

\subsection{Qualitative Comparison}
\begin{figure}[t]
\centering
    \includegraphics[width=0.5\textwidth]{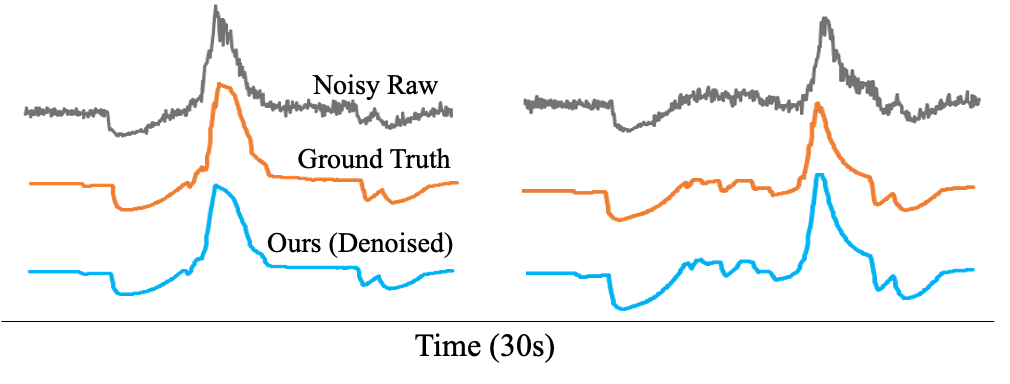}
\vspace{-10pt}
\caption{{Comparative Traces of Two Example ROIs from the simulated calcium imaging.}}
% Traces of two example ROIs from the simulated calcium imaging from the noisy recording, clean ground truth and our denoised output. }}
\label{fig:traces}
\end{figure}

The visual comparison of our model against other models on microscopy datasets as shown in Figures \ref{fig:teaser}, \ref{fig:visual_cmp}, and \ref{fig:visual_2pcal}, highlights the denoising effectiveness on two-photon calcium imaging. Our model demonstrates an exceptional ability to remove noise while retaining crucial signal details in calcium imaging. A detailed inspection in Figure \ref{fig:visual_cmp} reveals that our technique successfully recovers intrinsic high-frequency information, often lost or overly smoothed in other methods. Similarly, the comparison results of one-photon imaging in Figure \ref{fig:musc} show our model's proficiency in denoising and preserving finer details. Additionally, Figure \ref{fig:visual_floro} illustrates the superiority of our approach in denoising fluorescence microscopy datasets over other supervised and unsupervised techniques, further affirming our model's robust denoising capabilities across various scenarios. Figure \ref{fig:traces} shows the extracted traces of two region-of-interests (ROIs) manually drawn from the simulated two-photon calcium imaging.

% \vspace{-10pt}

\newcommand{\metric}{\multicolumn{1}{c}{PSNR} & \multicolumn{1}{c}{SSIM}}
\subsection{Ablation Study}

We conduct an ablation study to explore the influence of various components on our model's effectiveness. 

% \noindent
\paragraph{Influence of the DeepTemporal Interpolation.}
This study probes the critical contribution of the Temporal Filter (TF) in our framework. In the experiment, we feed the output from the feature generator straight to the Denoiser and evaluate the denoising performance using the variant without TF, denoting as $G_\phi$ + $D_\theta$. The results presented in Table \ref{tab:TFfilter} underline that the TF plays a substantial role in enhancing denoising efficiency.

\begin{table}
% \resizebox{\textwidth}{}{%
% \fontsize{3pt}{3pt}{3pt}{3pt}{3pt}{3pt}{3pt}{3pt}{3pt}{3pt}{3pt}\selectfont
\small
  \centering
  \setlength{\abovetopsep}{7pt}
  \setlength{\tabcolsep}{3.2pt}
  \renewcommand{\arraystretch}{1}
  \begin{tabular}{*{7}{l}}
    \toprule
    & \multicolumn{2}{c}{Gaussian 30} & \multicolumn{2}{c}{Poisson 30}   & \multicolumn{2}{c}{Impulse 0.2}  \\
    \cmidrule(lr){2-3}
    \cmidrule(lr){4-5}
    \cmidrule(lr){6-7}
 & \metric & \metric  &\metric \\
    \midrule
    $G_\phi$ + $D_\theta$ &  21.23 & 0.42 & 21.26 & 0.49 & 13.88 & 0.16 \\[1pt]
    $G_\phi$ + \textit{TF} + $D_\theta$ &  \textbf{31.67} & \textbf{0.90}& \textbf{30.82} & \textbf{0.92 }& \textbf{24.32} & \textbf{0.82} \\
    \bottomrule
  \end{tabular}
  \vspace{-8pt}

    % Without TF &  21.23/0.42 & 16.85/0.26 & 12.93/0.16 & 21.26/0.49 & 18.91/0.40 & 16.24/0.31 & 13.88/0.16 & 12.20/0.12 & 11.08/0.09 \\[1pt]
    % With TF &  31.28/0.89 & 28.07/0.84 & 22.08/0.72 & 31.04/0.93 & 28.92/0.90 & 25.24/0.88 & 23.58/0.79 & 20.42/0.72 & 18.40/0.64  \\
  
\caption{\textbf{Effect of Excluding Temporal Filter (TF) on Denoising Effectiveness.} This table presents a contrast in PSNR/SSIM metrics for our LIVE-YT-HFR-trained model, comparing denoising outcomes with and without including the temporal filter layer.}
\label{tab:TFfilter}
\vspace{-4pt}
\end{table}

% \noindent
\paragraph{Impact of Frame Rates Variation.}
In our study, we systematically examined the performance of our model across various frame rates, focusing primarily on high-speed videos. We subjected the LIVE-YT-HFR dataset to varying frame rates, including 120, 98, 82, 60, 30, and 24 fps, while introducing different noise types and intensities. Table \ref{tab:FPSablation} provides a detailed breakdown of our findings.

Notably, as evidenced by the results, our model showcases consistently superior denoising performance at higher frame rates, particularly at 120 fps across all noise types. However, the performance starts to wane with lower frame rate spectrums, especially at 30 and 24 fps. This highlights a potential limitation of our model regarding denoising videos with lower frame rates. This degradation in performance at lower frame rates suggests a possible direction for future improvements and optimizations to our algorithm.

\begin{table}
% \resizebox{\textwidth}{}{%
% \fontsize{3pt}{3pt}{3pt}{3pt}{3pt}{3pt}{3pt}{3pt}{3pt}{3pt}{3pt}\selectfont
\small
  \centering
  \setlength{\abovetopsep}{1pt}
  \setlength{\tabcolsep}{4.5pt}
  \renewcommand{\arraystretch}{1}
  \begin{tabular}{*{7}{l}}
    \toprule
    & \multicolumn{2}{c}{Gaussian 30} & \multicolumn{2}{c}{Poisson 30}   & \multicolumn{2}{c}{Impulse 0.2}  \\
    \cmidrule(lr){2-3}
    \cmidrule(lr){4-5}
    \cmidrule(lr){6-7}
 $fps$ & \metric & \metric  &\metric \\
    \midrule
    120 &  \textbf{31.67} & \textbf{0.90} & \textbf{30.82} & \textbf{0.92} & \textbf{24.32} & \textbf{0.82} \\[1pt]
    98 &  28.83 & 0.83  & 28.69 & 0.88  & 22.44 & 0.73  \\[1pt]
    82 &  \underline{29.30} & \underline{0.85 } & \underline{29.42} & \underline{0.89 } & 22.67 & 0.74  \\[1pt]
    60 &  29.10 & \underline{0.85 } & 28.90 & \underline{0.89 } & \underline{22.91} & \underline{0.75 }  \\[1pt]
    30 &  24.29 & 0.71  & 24.68 & 0.76  & 21.25 & 0.65  \\[1pt]
    24 &  23.58 & 0.68  & 23.50 & 0.72  & 20.43 & 0.59   \\
    \bottomrule
  \end{tabular}
  \vspace{-8pt}
  \caption{\textbf{Impact of Frame Rate Variation.} This table showcases the PSNR/SSIM metrics across various frame rates from the LIVE-YT-HFR dataset.}
\label{tab:FPSablation}
\end{table}

% Through these ablation studies, we gained insights into the critical role of these parameters in our model's denoising performance, which could serve as guidance for future work in this area.

\paragraph{Different of Number of Input Frames (N)}
We conducted an ablation study on the number of frames ($N$) utilized as the input batch in our model. The findings in Table \ref{tab:frames} indicate that $N = 7$ yields the optimal performance across different noise types.

\begin{table}[h]
\small
  \centering
  \setlength{\abovetopsep}{7pt}
  \setlength{\tabcolsep}{6pt}
  \renewcommand{\arraystretch}{1.2}
  \begin{tabular}{ccccccc}
  \hline
  \centering
   &  & N=$3$ & N=$5$ & N=$7$ & N=$9$ & N=$11$\\
   \hline
   % & Gaussian $50$ & {27.87} & {28.51} & {28.70} & {} & {}\\
   & Poisson $50$ & \underline{27.35} & {25.81} & \textbf{28.72} & {26.97} & {26.23}\\
   & Impulse $0.2$ & {23.93} & \underline{24.42} & \textbf{24.44} & {24.07} & {24.22}\\
   \hline
  \end{tabular}
    \vspace{-5pt}
  \caption{\textbf{Impact of Number of Input Frames.} This table displays the average PSNR values for various numbers of input frames from the LIVE-YT-HFR dataset.}
  \label{tab:frames}
\end{table}

\section{Discussion}

% This paper presents a novel SpatioTemporal Weighted (STW) kernel-based approach for robust high frame rate video denoising under high-intensity noise, outperforming prevalent unsupervised networks like UDVD and DeepInterpolation. The success of our method across varied noise types and intensities is evident in our extensive evaluations.
% 
% The ablation study reinforces the impact of STW shape on the final output quality, suggesting a need for application-specific parameter tuning. However, our current model struggles with videos featuring fast-moving objects, leading to missing sections or blurry outputs, thus calling for future enhancements.
We presented a novel unsupervised approach to denoise microscopy videos, leveraging the DeepTemporal Interpolation based CNN network architecture.
We demonstrate that our unsupervised microsopy video denoising method achieves outperforming performance on various real-world micoscopy dataset and diverse noise types by emphasizing the importance of spatiotemporal features in microscopy videos. 
In our future work, we plan to enhance our model's efficiency with low frame rate videos. Additionally, we intend to refine our temporal filter to adjust the parameters dynamically based on observed motion speed. We also aim to adaptive accommodate the number of input frames in the denoisng batch according to the speed of moving objects in the videos.

% The systematic exploration of the frame rate's influence on denoising results further deepened our understanding, highlighting both the strengths and potential areas for improvement in our approach.
% 
% One solution could be to integrate object detection or optical flow estimation to dynamically adjust temop parameter, thus better accommodating fast movements. 
% Furthermore, future work should seek to optimize our relatively time-consuming training process, possibly through more efficient algorithms or leveraging advanced computational resources.
% 
% In the future, we aim to extend our approach's capabilities to work efficiently with low frame rate videos and those featuring fast-moving objects. The endeavor will involve tackling the challenges we currently face, adopting innovative solutions, and refining our model's architecture.

%%%%%%%%% REFERENCES
{\small
\bibliographystyle{ieee_fullname}
\bibliography{egbib}

\begin{thebibliography}{10}\itemsep=-1pt

\bibitem{al2020critical}
Zainab~T Al-Sharify, Talib~A Al-Sharify, Noor~T Al-Sharify, et~al.
\newblock A critical review on medical imaging techniques (ct and pet scans) in the medical field.
\newblock In {\em IOP Conference Series: Materials Science and Engineering}, volume 870, page 012043. IOP Publishing, 2020.

\bibitem{noise2self}
Joshua Batson and Loic Royer.
\newblock Noise2self: Blind denoising by self-supervision.
\newblock In {\em International Conference on Machine Learning}, pages 524--533. PMLR, 2019.

\bibitem{berdyyeva2014zolpidem}
Tamara Berdyyeva, Stephani Otte, Leah Aluisio, Yaniv Ziv, Laurie~D Burns, Christine Dugovic, Sujin Yun, Kunal~K Ghosh, Mark~J Schnitzer, Timothy Lovenberg, et~al.
\newblock Zolpidem reduces hippocampal neuronal activity in freely behaving mice: a large scale calcium imaging study with miniaturized fluorescence microscope.
\newblock {\em PLoS One}, 9(11):e112068, 2014.

\bibitem{burgess2016hunger}
Christian~R Burgess, Rohan~N Ramesh, Arthur~U Sugden, Kirsten~M Levandowski, Margaret~A Minnig, Henning Fenselau, Bradford~B Lowell, and Mark~L Andermann.
\newblock Hunger-dependent enhancement of food cue responses in mouse postrhinal cortex and lateral amygdala.
\newblock {\em Neuron}, 91(5):1154--1169, 2016.

\bibitem{videnn}
Michele Claus and Jan Van~Gemert.
\newblock Videnn: Deep blind video denoising.
\newblock In {\em Proceedings of the IEEE/CVF conference on computer vision and pattern recognition workshops}, pages 0--0, 2019.

\bibitem{mf2f}
Val{\'e}ry Dewil, J{\'e}r{\'e}my Anger, Axel Davy, Thibaud Ehret, Gabriele Facciolo, and Pablo Arias.
\newblock Self-supervised training for blind multi-frame video denoising.
\newblock In {\em Proceedings of the IEEE/CVF winter conference on applications of computer vision}, pages 2724--2734, 2021.

\bibitem{frame2frame}
Thibaud Ehret, Axel Davy, Jean-Michel Morel, Gabriele Facciolo, and Pablo Arias.
\newblock Model-blind video denoising via frame-to-frame training.
\newblock In {\em Proceedings of the IEEE/CVF conference on computer vision and pattern recognition}, pages 11369--11378, 2019.

\bibitem{hussain2022modern}
Shah Hussain, Iqra Mubeen, Niamat Ullah, Syed Shahab Ud~Din Shah, Bakhtawar~Abduljalil Khan, Muhammad Zahoor, Riaz Ullah, Farhat~Ali Khan, Mujeeb~A Sultan, et~al.
\newblock Modern diagnostic imaging technique applications and risk factors in the medical field: a review.
\newblock {\em BioMed research international}, 2022, 2022.

\bibitem{kasban2015comparative}
Hany Kasban, MAM El-Bendary, and DH Salama.
\newblock A comparative study of medical imaging techniques.
\newblock {\em International Journal of Information Science and Intelligent System}, 4(2):37--58, 2015.

\bibitem{noise2void}
Alexander Krull, Tim-Oliver Buchholz, and Florian Jug.
\newblock Noise2void-learning denoising from single noisy images.
\newblock In {\em Proceedings of the IEEE/CVF conference on computer vision and pattern recognition}, pages 2129--2137, 2019.

\bibitem{self-supervised}
Samuli Laine, Tero Karras, Jaakko Lehtinen, and Timo Aila.
\newblock High-quality self-supervised deep image denoising.
\newblock {\em Advances in Neural Information Processing Systems}, 32, 2019.

\bibitem{deepinterpolation}
J{\'e}r{\^o}me Lecoq, Michael Oliver, Joshua~H Siegle, Natalia Orlova, Peter Ledochowitsch, and Christof Koch.
\newblock Removing independent noise in systems neuroscience data using deepinterpolation.
\newblock {\em Nature methods}, 18(11):1401--1408, 2021.

\bibitem{noise2noise}
Jaakko Lehtinen, Jacob Munkberg, Jon Hasselgren, Samuli Laine, Tero Karras, Miika Aittala, and Timo Aila.
\newblock Noise2noise: Learning image restoration without clean data.
\newblock {\em arXiv preprint arXiv:1803.04189}, 2018.

\bibitem{vrt}
Jingyun Liang, Jiezhang Cao, Yuchen Fan, Kai Zhang, Rakesh Ranjan, Yawei Li, Radu Timofte, and Luc Van~Gool.
\newblock Vrt: A video restoration transformer.
\newblock {\em arXiv preprint arXiv:2201.12288}, 2022.

\bibitem{rvrt}
Jingyun Liang, Yuchen Fan, Xiaoyu Xiang, Rakesh Ranjan, Eddy Ilg, Simon Green, Jiezhang Cao, Kai Zhang, Radu Timofte, and Luc~V Gool.
\newblock Recurrent video restoration transformer with guided deformable attention.
\newblock {\em Advances in Neural Information Processing Systems}, 35:378--393, 2022.

\bibitem{deformableattention}
Jingyun Liang, Yuchen Fan, Xiaoyu Xiang, Rakesh Ranjan, Eddy Ilg, Simon Green, Jiezhang Cao, Kai Zhang, Radu Timofte, and Luc~V Gool.
\newblock Recurrent video restoration transformer with guided deformable attention.
\newblock {\em Advances in Neural Information Processing Systems}, 35:378--393, 2022.

\bibitem{lindner2023lightweight}
Lydia Lindner, Alexander Effland, Filip Ilic, Thomas Pock, and Erich Kobler.
\newblock Lightweight video denoising using aggregated shifted window attention.
\newblock In {\em Proceedings of the IEEE/CVF Winter Conference on Applications of Computer Vision}, pages 351--360, 2023.

\bibitem{livehfr}
Pavan~C. Madhusudana, Xiangxu Yu, Neil Birkbeck, Yilin Wang, Balu Adsumilli, and Alan~C. Bovik.
\newblock Subjective and objective quality assessment of high frame rate videos.
\newblock {\em IEEE Access}, 9:108069--108082, 2021.

\bibitem{mao2016image}
Xiaojiao Mao, Chunhua Shen, and Yu-Bin Yang.
\newblock Image restoration using very deep convolutional encoder-decoder networks with symmetric skip connections.
\newblock {\em Advances in neural information processing systems}, 29, 2016.

\bibitem{pytorch}
Adam Paszke, Sam Gross, Francisco Massa, Adam Lerer, James Bradbury, Gregory Chanan, Trevor Killeen, Zeming Lin, Natalia Gimelshein, Luca Antiga, et~al.
\newblock Pytorch: An imperative style, high-performance deep learning library.
\newblock {\em Advances in neural information processing systems}, 32, 2019.

\bibitem{pinto2015cell}
Lucas Pinto and Yang Dan.
\newblock Cell-type-specific activity in prefrontal cortex during goal-directed behavior.
\newblock {\em Neuron}, 87(2):437--450, 2015.

\bibitem{unet}
Olaf Ronneberger, Philipp Fischer, and Thomas Brox.
\newblock U-net: Convolutional networks for biomedical image segmentation.
\newblock In {\em Medical Image Computing and Computer-Assisted Intervention--MICCAI 2015: 18th International Conference, Munich, Germany, October 5-9, 2015, Proceedings, Part III 18}, pages 234--241. Springer, 2015.

\bibitem{sawinski2009visually}
Juergen Sawinski, Damian~J Wallace, David~S Greenberg, Silvie Grossmann, Winfried Denk, and Jason~ND Kerr.
\newblock Visually evoked activity in cortical cells imaged in freely moving animals.
\newblock {\em Proceedings of the National Academy of Sciences}, 106(46):19557--19562, 2009.

\bibitem{udvd}
Dev~Yashpal Sheth, Sreyas Mohan, Joshua~L Vincent, Ramon Manzorro, Peter~A Crozier, Mitesh~M Khapra, Eero~P Simoncelli, and Carlos Fernandez-Granda.
\newblock Unsupervised deep video denoising.
\newblock In {\em Proceedings of the IEEE/CVF International Conference on Computer Vision}, pages 1759--1768, 2021.

\bibitem{naomi}
Alexander Song, Jeff~L Gauthier, Jonathan~W Pillow, David~W Tank, and Adam~S Charles.
\newblock Neural anatomy and optical microscopy (naomi) simulation for evaluating calcium imaging methods.
\newblock {\em Journal of neuroscience methods}, 358:109173, 2021.

\bibitem{song2022tempformer}
Mingyang Song, Yang Zhang, and Tun{\c{c}}~O Ayd{\i}n.
\newblock Tempformer: Temporally consistent transformer for video denoising.
\newblock In {\em European Conference on Computer Vision}, pages 481--496. Springer, 2022.

\bibitem{dvdnet}
Matias Tassano, Julie Delon, and Thomas Veit.
\newblock Dvdnet: A fast network for deep video denoising.
\newblock In {\em 2019 IEEE International Conference on Image Processing (ICIP)}, pages 1805--1809. IEEE, 2019.

\bibitem{fastdvdnet}
Matias Tassano, Julie Delon, and Thomas Veit.
\newblock Fastdvdnet: Towards real-time deep video denoising without flow estimation.
\newblock In {\em Proceedings of the IEEE/CVF conference on computer vision and pattern recognition}, pages 1354--1363, 2020.

\bibitem{ulman2017objective}
Vladim{\'\i}r Ulman, Martin Ma{\v{s}}ka, Klas~EG Magnusson, Olaf Ronneberger, Carsten Haubold, Nathalie Harder, Pavel Matula, Petr Matula, David Svoboda, Miroslav Radojevic, et~al.
\newblock An objective comparison of cell-tracking algorithms.
\newblock {\em Nature methods}, 14(12):1141--1152, 2017.

\bibitem{wang2023recurrent}
Zichun Wang, Yulun Zhang, Debing Zhang, and Ying Fu.
\newblock Recurrent self-supervised video denoising with denser receptive field.
\newblock In {\em Proceedings of the 31st ACM International Conference on Multimedia}, pages 7363--7372, 2023.

\bibitem{rvidenet}
Huanjing Yue, Cong Cao, Lei Liao, Ronghe Chu, and Jingyu Yang.
\newblock Supervised raw video denoising with a benchmark dataset on dynamic scenes.
\newblock In {\em Proceedings of the IEEE/CVF conference on computer vision and pattern recognition}, pages 2301--2310, 2020.

\bibitem{dncnn}
Kai Zhang, Wangmeng Zuo, Yunjin Chen, Deyu Meng, and Lei Zhang.
\newblock Beyond a gaussian denoiser: Residual learning of deep cnn for image denoising.
\newblock {\em IEEE transactions on image processing}, 26(7):3142--3155, 2017.

\bibitem{RCD}
Zhaoyang Zhang and et al.
\newblock Real-time controllable denoising for image and video.
\newblock In {\em CVPR}, pages 14028--14038, 2023.

\bibitem{untrainednet}
Huan Zheng, Tongyao Pang, and Hui Ji.
\newblock Unsupervised deep video denoising with untrained network.
\newblock In {\em Proceedings of the AAAI Conference on Artificial Intelligence}, volume~37, pages 3651--3659, 2023.

\bibitem{ziv2013long}
Yaniv Ziv, Laurie~D Burns, Eric~D Cocker, Elizabeth~O Hamel, Kunal~K Ghosh, Lacey~J Kitch, Abbas~El Gamal, and Mark~J Schnitzer.
\newblock Long-term dynamics of ca1 hippocampal place codes.
\newblock {\em Nature neuroscience}, 16(3):264--266, 2013.

\end{thebibliography}
}

\end{document}